\documentclass[twocolumn]{aastex701}

\usepackage{physics}
\usepackage{bm}
\newcommand{\msun}{{\rm M}_\odot}

\begin{document}

\title{An Origin of Radially Aligned Filaments in Hub-Filament Systems}

\correspondingauthor{Shingo Nozaki}
\author[orcid=0000-0003-4271-4901]{Shingo Nozaki}
\affiliation{Department of Earth and Planetary Sciences, Faculty of Science, Kyushu University, Nishi-ku, Fukuoka 819-0395, Japan}
\email[show]{nozaki.shingo.307@s.kyushu-u.ac.jp}

\author[orcid=0000-0003-4366-6518]{Shu-ichiro Inutsuka} 
\affiliation{Department of Physics, Graduate School of Science, Nagoya University, Furo-cho, Chikusa-ku, Nagoya, Aichi 464-8602, Japan}
\email{inutsuka.shu-ichiro.i2@f.mail.nagoya-u.ac.jp}

\begin{abstract}
Recent observations have identified hub-filament systems (HFSs) as the primary formation sites of massive stars and star clusters. Some HFSs are characterized by multiple filaments aligned radially toward a central high-density hub. However, the physical origin of radially aligned filaments remains unknown. Here, we propose a new formation mechanism of HFSs driven by the interaction of a fast magnetohydrodynamic shock with a molecular cloud characterized by an hourglass-shaped magnetic field and density inhomogeneity. Our three-dimensional magnetohydrodynamic simulations show that the shock propagation leads to the formation of radially aligned filamentary structures with line masses slightly above the thermally critical line mass and lengths of $1\text{--}3\,\mathrm{pc}$, and widths of $0.06\text{–}0.08\,\mathrm{pc}$. High-density filamentary gas ($n_{\mathrm{H_2}}\sim10^4\,\mathrm{cm^{-3}}$) selectively exhibits inward velocities of $1\text{--}4\,\mathrm{km\,s^{-1}}$ that increase toward the hub center, while the ambient low-density inter-filament gas retains low velocities regardless of the radius. Mass accretion onto the hub is channeled through dense filaments. The filament formation is driven by oblique shocks generated at the bent magnetic field lines. The resulting post-shock amplification of the tangential magnetic field induces a magnetically guided inflow. The shock-interface interaction amplifies density perturbations, resembling Richtmyer--Meshkov instability modes, which promotes the fragmentation of the shocked layer into multiple filaments. The process studied in this Letter explains both the morphology of radially aligned filaments and the selective mass accretion observed in HFSs. In our simulation, the resulting star formation efficiency is $\sim4\%$, suggesting that the shock-driven evolution limits the SFE to only a few percent.
\end{abstract}


\section{Introduction}
Filamentary structures in molecular clouds are the primary sites of star formation. Among filamentary structures, the hub-filament system (HFS), consisting of filaments radially converging to a central dense hub, is a key site for cluster and high-mass star formation \citep[e.g.,][]{myers2009,kumar2020}. Observations using Herschel, Spitzer, and ALMA have revealed such parsec-scale systems in various star-forming regions. For example, in the Mon~R2 cloud, a network of filaments aligns radially toward the hub center \citep{kumar2022}. Similar structures are identified in both high- and low-mass star-forming regions \citep[e.g.,][]{dewangan2025}, suggesting their universality in clustered star formation. While magnetic fields and gas kinematics in HFSs have been observationally characterized \citep[e.g.,][]{wang2019,wang2020a,wang2020b,hwang2022,zhou2023}, the physical origin of the HFS morphology remains poorly understood.

Theoretical studies have proposed various mechanisms for filament formation. \citet{inoue2018} showed, using simulations of shock-cloud interactions, that a shock propagating perpendicular to the magnetic field generates complex filamentary structures via the formation of oblique shocks. \citet{abe2021} revealed that formation mechanisms can be classified into four types based on the field-flow orientation, indicating that the interaction between the shock and the magnetic field plays a crucial role in filament formation. On the other hand, as attempts to understand the HFS formation process, numerical simulations focusing on individual filament-filament collisions \citep[e.g.,][]{kashiwagi2024} and gravitational structure formation models based on global hierarchical collapse \citep[e.g.,][]{vazquez-semadeni2025} have been proposed. However, the formation of distinct radially aligned filaments converging to a central hub, as observed in HFSs, has not yet been reproduced. While previous shock-cloud interaction studies assumed a shock propagating perpendicular to a uniform magnetic field, the effect of field curvature has not been considered. Given that molecular clouds are threaded by hourglass-shaped magnetic fields due to self-gravity, investigating the interaction between a shock and such curved field geometries is crucial for understanding the origin of HFSs.

In this Letter, we investigate the interaction of a fast-mode shock with a molecular cloud characterized by an hourglass-shaped magnetic field and density inhomogeneity, using three-dimensional ideal magnetohydrodynamic (MHD) simulations. We demonstrate that the shock-cloud interaction generates multiple filaments radially aligned toward a central hub. Based on these results, we propose a new scenario for HFSs driven by external shocks, such as those from supernova remnants or expanding H II regions, interacting with the pre-existing curved magnetic field.

\section{Numerical Settings} 
To investigate the formation mechanism of the HFSs, we have performed three-dimensional ideal MHD simulations using the adaptive mesh refinement code SFUMATO \citep{matsumoto2007,matsumoto2015,fukushima2021,nozaki2025}, which solves the equations of ideal MHD with self-gravity \citep[see also][for the basic equations]{matsumoto2007}. The heating and cooling processes include gas–dust energy exchange, line emission, and chemical heating, implemented with a simplified chemical network \citep{fukushima2020,fukushima2021}. To model the interaction between a molecular cloud and an external shock originating from a supernova remnant, we consider a cloud placed within a cube simulation box with a side length of $10\,\mathrm{pc}$. The background number density in the computational domain is initially set to $n_{\mathrm{bg}}=1.01\times10^{2}\,\mathrm{cm^{-3}}$, comparable to the typical density of a giant molecular cloud envelope \citep[e.g.,][]{heyer2015}. We adopt the initial condition of a cloud that is flattened along the $z$-direction. The density distribution smoothly connects to the background and is given by
\begin{equation}
n_{\mathrm{H_2}}=n_{\mathrm{bg}}+n_{\mathrm{bg}}\,(A_{\mathrm{max}}-1)
\left[1+\frac{x^2+y^2+((z-z_{\mathrm{off}})/\alpha)^2}{R_{\mathrm{flat}}^2}\right]^{-2},
\end{equation}
where $A_{\mathrm{max}}$ is the peak-to-background density contrast, set to $A_{\mathrm{max}}=130$. The parameters $\alpha$ and $R_{\mathrm{flat}}$ characterize the degree of vertical flattening and the radius of the central flat region, set to $\alpha=0.1$ and $R_{\mathrm{flat}}=1.75\,\mathrm{pc}$, respectively. The vertical offset of the cloud center is set to $z_{\mathrm{off}}=+1.45\,\mathrm{pc}$. The total gas mass in the computational domain is $8.6\times10^3\,\msun$. The mass of the initial cloud, defined as gas with $n_{\mathrm{H_2}}>1.02\times10^2\,\mathrm{cm^{-3}}$, is $4.8\times10^3\,\msun$. The initial magnetic field is uniform and oriented along the $z$-axis with a strength of $B_{\mathrm{0}}=15\,\mu\mathrm{G}$, consistent with typical values observed in nearby molecular clouds \citep{crutcher2010,pattle2023}. The corresponding mass-to-flux ratio of the entire computational domain is $1.9$, indicating that the cloud is slightly supercritical. The base grid consists of $128^3$ cells. At the finest refinement level ($l_{\mathrm{max}}=4$), the cell width is $\Delta x=9.8\times10^{-3}\,\mathrm{pc}$, which resolves the local Jeans length by more than five cells. Sink particles are introduced when the gas density exceeds $n_{\mathrm{H_2, thr}}=1\times10^{5}\,\mathrm{cm^{-3}}$ in order to follow the long-term evolution of the cloud.

A planar shock propagating in the negative $z$-direction is injected from the upper boundary ($z=+5\,\mathrm{pc}$) of the computational domain. As fluid boundary conditions, the $z=+5\,\mathrm{pc}$ plane is treated as a fixed inflow boundary, while the remaining five boundaries are set to free boundaries. At the fixed boundary, an inflow velocity of $v=(0,0,v_{\mathrm{shock}})$ is imposed, with $v_{\mathrm{shock}}=-5.0\,\mathrm{km\,s^{-1}}$. All other physical quantities at this boundary are fixed to their initial ambient values outside the cloud, except for the gas pressure, enhanced to $5\times10^2$ times the ambient pressure in order to drive the shock. This inflow velocity corresponds to sonic and background Alfv\'enic Mach numbers of approximately $27$ and $2.3$, respectively, for a $10\,\mathrm{K}$ molecular gas, exceeding the fast-magnetosonic speed. Following \citet{matsumoto2015}, we initialize a turbulent velocity field throughout the computational domain with a root-mean-square Mach number of $\mathcal{M}_{\mathrm{rms}}=2$, allowed to decay without continuous driving.

Before the shock reaches the molecular cloud, the initially uniform magnetic field aligned with the $z$-axis becomes slightly pinched near the cloud center due to gravitational contraction, forming a weak hourglass-shaped magnetic field morphology. Such hourglass-shaped magnetic field structures on molecular cloud scales have been reported in observations of massive star-forming regions \citep[e.g.,][]{beltran2019}. In the fiducial ($\psi=0^\circ$) configuration, the fast-mode shock propagates nearly parallel to the large-scale magnetic field along the $z$-direction. The initially imposed turbulent velocity field introduces weak inhomogeneity in the density distribution of the molecular cloud prior to the shock impact. The simulation is evolved up to $0.5\,\mathrm{Myr}$ after the shock reaches the cloud, which is sufficient to capture the formation of the HFS and its early dynamical evolution. Extending the calculation to later times would require prohibitively small time steps, mainly due to the high Alfvén speeds and outflow velocities that are generated by the shock--cloud interaction. This would lead to a substantial increase in computational cost.

In addition to the fiducial configuration ($\psi=0^\circ$) adopted as the reference model, we performed supplementary simulations in which the magnetized cloud system is rotated as a whole, while the shock propagation direction is kept fixed along the $z$-axis. This setup introduces relative inclination angles of $\psi=10^\circ,\,15^\circ,\,20^\circ$ and $30^\circ$ between the shock direction and the magnetic field, compared to the fiducial $\psi=0^\circ$ case.

In this Letter, we focus on the $\psi=0^\circ$ model as a representative case to demonstrate the key physical processes, specifically at the epoch when the HFS is clearly established, while showing the $\psi=15^\circ$ and $30^\circ$ models to assess the robustness of the morphology against deviations from the idealized geometry. A systematic parameter survey is deferred to future work.

\section{Results}
\subsection{Formation of Radially Aligned Filaments}
\begin{figure*}[htbp]
    \centering
    \includegraphics[width=\textwidth]{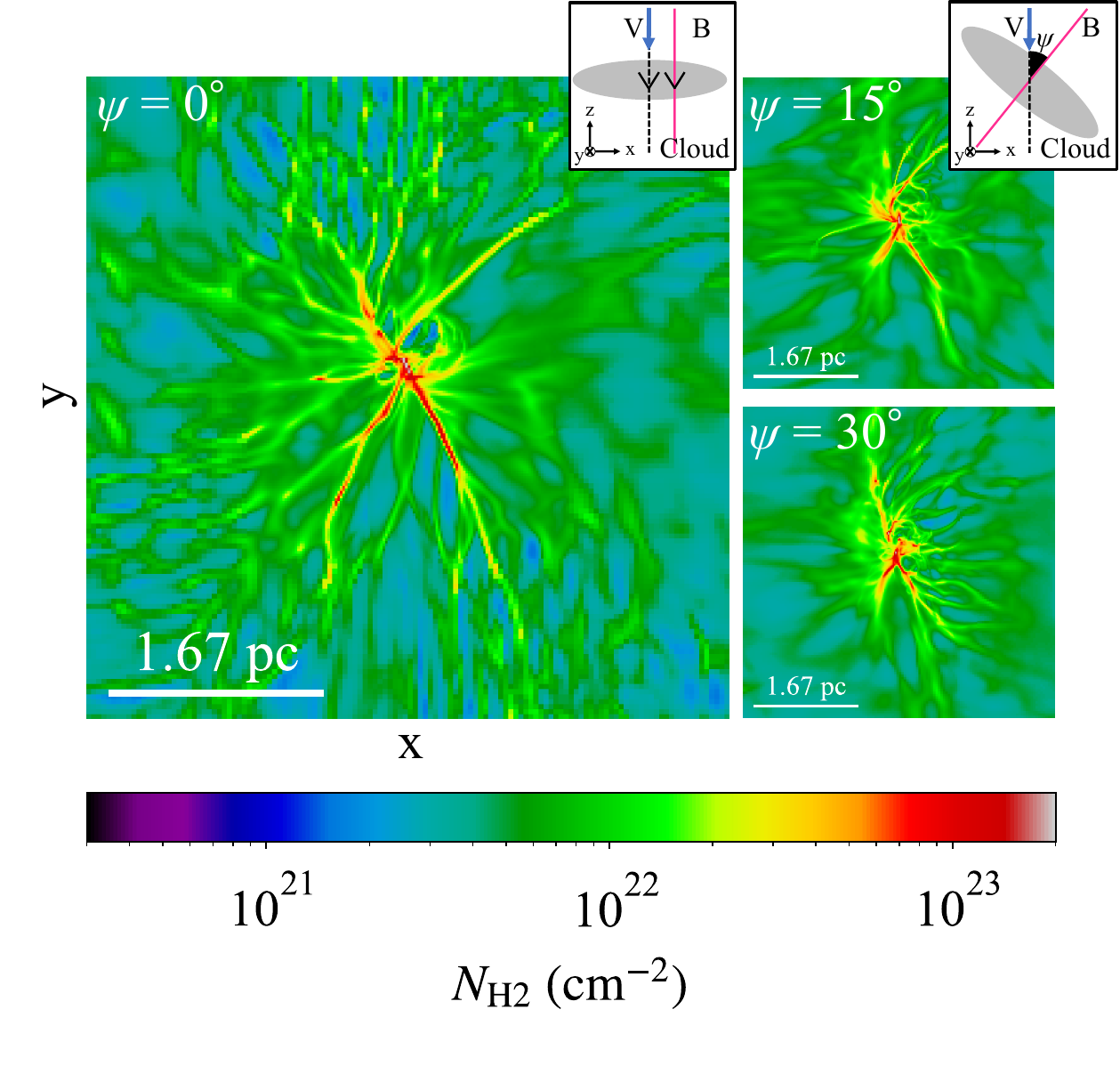} 
    \caption{$\mathrm{H_2}$ column density maps in the $x$--$y$ plane at $t=0.5\,\mathrm{Myr}$ after the shock sweeps the cloud for different inclination angles $\psi$ between the shock propagation direction and the magnetic-field axis. The panels show $\psi=0^\circ$, $15^\circ$, and $30^\circ$. The color scale represents the $\mathrm{H_2}$ column density integrated along the $z$-axis. The box size is $5.0\,\mathrm{pc}$ on each side. Small inset schematics in each panel illustrate the definition of the inclination angle $\psi$ (between the shock propagation direction and the magnetic-field axis) and the corresponding initial cloud--field geometry; the gray shape, magenta line, and blue arrow indicate the initial cloud density profile, magnetic-field direction, and shock propagation direction, respectively.}
    \label{hf}
\end{figure*}

As shown in the left panel of Figure~\ref{hf}, a HFS develops when a fast-mode shock reaches the central region of the cloud with an hourglass-shaped magnetic field and weak density inhomogeneity introduced by turbulence.
Multiple filaments with lengths of $1\text{--}3\,\mathrm{pc}$ extend radially from the central hub, exhibiting a spoke-like morphology. Their radial configuration and column densities are comparable to those observed in HFSs \citep[e.g.,][]{kumar2022,dewangan2025}.

We also show simulations in which the shock propagation direction is inclined by $\psi=15^\circ$ and $30^\circ$ relative to the magnetic-field axis (Figure~\ref{hf}, right panels). In both cases, radially aligned filaments develop around the central hub. Although the overall symmetry decreases with increasing inclination angle, the tendency for filament alignment toward the hub is preserved. In the $\psi=30^\circ$ case, filament formation tends to be weaker on the one side of the cloud that is struck first by the shock (the left side in $\psi=30^\circ$ panel of Figure~\ref{hf}), leading to a more asymmetric hub-filament morphology. This likely reflects the fact that the shock interacts with the cloud and the curved magnetic-field structure at different times and with different effective obliquities across the cloud. In the following analysis, we focus on the $\psi=0^\circ$ case, which provides the clearest illustration of the underlying mechanism.

\begin{figure}[htbp]
    \centering
    \includegraphics[width=\columnwidth]{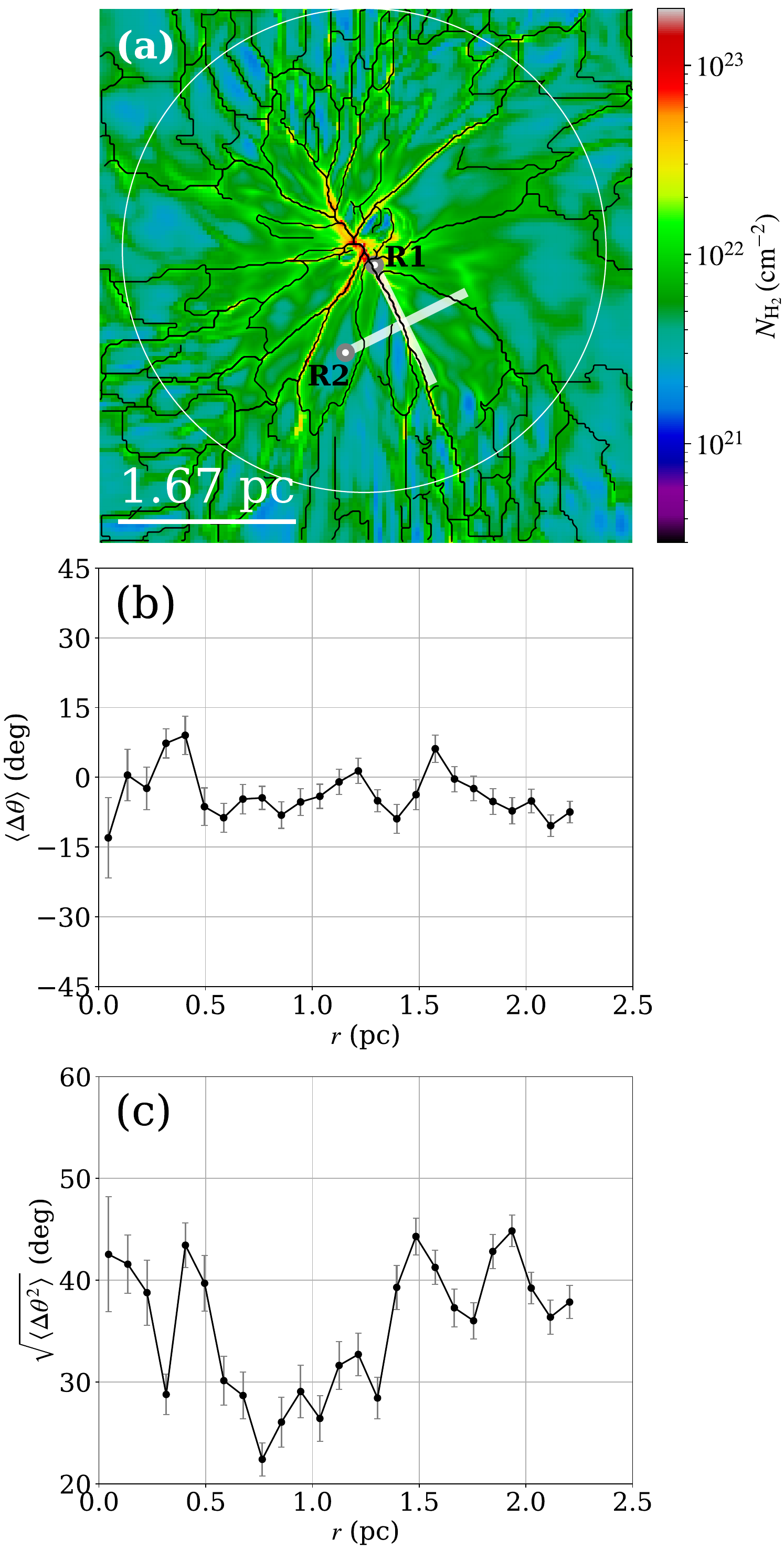}
    \caption{(a) $\mathrm{H_2}$ column density map from the left panel of Figure~\ref{hf}, overlaid with filament skeletons identified using \textsc{DisPerSE} \citep{sousbie2011a,sousbie2011b}. The lines labeled R1 and R2 denote cuts parallel and perpendicular to the filaments, respectively (see Figure~\ref{v_b} for details). The white circle (diameter of $4.5\,\mathrm{pc}$) indicates the region centered on the density peak. (b)--(c) Radial profiles of the mean angle $\langle\Delta\theta\rangle$ (b) and the RMS angle (c) of the filaments relative to the radial vector from the hub center. The hub center is defined as the column density peak, distinct from the geometric center of the simulation box. The angle $\Delta\theta$ ranges from $-90^\circ$ to $90^\circ$, where $0^\circ$ corresponds to the radial direction. The plots cover the region within the white circle in (a). Error bars represent the standard errors.}
    \label{filfinder}
\end{figure}

We identified filamentary structures using the \textsc{DisPerSE} \citep{sousbie2011a,sousbie2011b}, as illustrated in Figure~\ref{filfinder}~(a).
The extraction was performed on logarithmic column density maps smoothed with a 2-pixel Gaussian kernel to suppress grid noise. We adopted a persistence threshold of $0.02\,\mathrm{dex}$ to detect faint structures (column density contrast $\gtrsim5\%$) while avoiding artifacts, and removed unphysical segments. We measured the characteristic width directly from the column density map. Transverse profiles were sampled perpendicular to the extracted filament skeletons and stacked to construct a mean radial profile. Fitting this profile with a Gaussian plus a constant background yields a characteristic full width at half maximum (FWHM) of $\sim0.07\,\mathrm{pc}$. Alternative estimators, including Plummer fitting with fixed $p=2$ and direct half-maximum measurements \citep[e.g.,][]{arzoumanian2019}, give consistent values in the range $0.06$\text{--}$0.08\,\mathrm{pc}$. Using this measured width, we defined the filament region as the area within one FWHM radius from the filament skeleton. The total filament mass was obtained by integrating the column density over the entire filament network identified with \textsc{DisPerSE}. We derived a mean $\mathrm{H_2}$ column density of $8\times10^{21}\,\mathrm{cm^{-2}}$ for the identified filaments. This value exceeds the star-formation threshold suggested by \citet{andre2010}, indicating that the identified filaments represent a dense filamentary network capable of hosting active star formation. The corresponding line mass is $22\,\msun\,\mathrm{pc^{-1}}$, slightly above the thermal critical line mass at $10\,\mathrm{K}$ 
($M_{\mathrm{line,cri}}\sim16.8\,\msun\,\mathrm{pc^{-1}}$), suggesting that these filaments are in a mildly supercritical state.

To quantify the geometric alignment, we measured the deviation angle, $\Delta\theta$, defined between the local tangent of each filament and the radial vector originating from the hub center as in \cite{kumar2022}. As shown in Figure~\ref{filfinder}~(b), the distribution of the mean deviation angle, $\langle\Delta\theta\rangle$, for filaments within a radius of $2.25\,\mathrm{pc}$ is tightly confined to the range of $-15^\circ$ to $15^\circ$, indicating that most filaments are closely aligned with the radial direction. As presented in Figure~\ref{filfinder}~(c), the root-mean-square (RMS) angles of the filaments relative to the radial vector fall within the range of $0^\circ\text{--}45^\circ$. These values are consistently lower than the dispersion expected for a random orientation ($\mathrm{RMS}\sim52^\circ$), confirming a significant tendency toward radial alignment. These statistical results quantitatively substantiate the formation of the radially aligned HFS visualized in the left panel of Figure~\ref{hf}, showing a remarkable agreement with \cite{kumar2022}.

\subsection{Velocity Structure of Hub-Filament System}
\begin{figure*}[htbp]
    \centering
    \includegraphics[width=\textwidth]{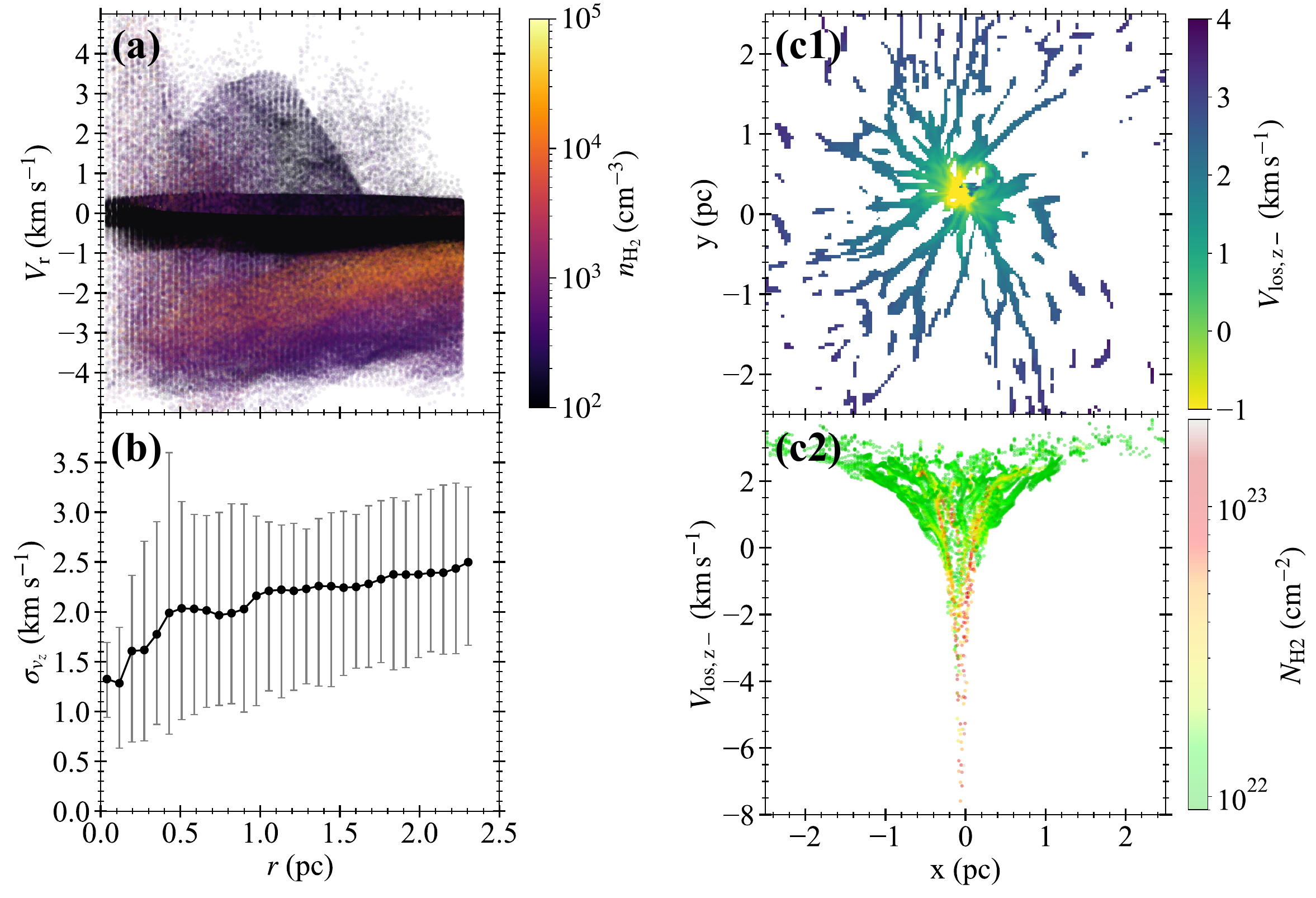}
    \caption{
    Velocity structure of the HFS $0.5\,\mathrm{Myr}$ after the shock impact. (a) Radial velocity $V_r$ plotted as a function of the cylindrical radius $r$ from the hub center. $V_r$ and $r$ are defined in a cylindrical coordinate system aligned with the $z$-axis. The color of each point corresponds to the $\mathrm{H_2}$ number density. Only cells within a cylindrical radius of $r=2.25\,\mathrm{pc}$ are plotted. (b) Median of the density-weighted velocity dispersion $\sigma_{v_z}$ along the $z$-axis as a function of $r$. Error bars indicate the $95\%$ confidence intervals. Note that in both panels, the hub center is defined as the position of the peak gas density at this epoch and does not coincide with the origin of the computational domain. (c) Observationally motivated kinematic maps constructed from the same snapshot as in the left panel of Figure~\ref{hf}. (c1) Column-density-selected line-of-sight velocity map. Only regions with $N_{\mathrm{H_2}} > 8\times10^{21}\,\mathrm{cm^{-2}}$, corresponding approximately to the mean column density of the identified filaments, are included. The line-of-sight velocity components are measured along the $z$-axis from an observer located at $z=-\infty$ and are shown in the laboratory frame. (c2) Position--velocity (PV) diagram showing the line-of-sight velocity $V_\mathrm{los, z-}$ as a function of the projected $x$-coordinate, weighted by the column density.}
    \label{velocity}
\end{figure*}

Figure~\ref{velocity}~(a) reveals a clear kinematic distinction between the dense and diffuse gas components.
The high-density gas ($n_{\mathrm{H_2}}\sim3.0\times10^4\,\mathrm{cm^{-3}}$, corresponding to filaments) exhibits high inward velocities ($V_r\sim-4$ to $-1\,\mathrm{km\,s^{-1}}$) that increase toward the hub center. In contrast, the ambient low-density gas retains low velocities ($V_r\sim-1$ to $+0.5\,\mathrm{km\,s^{-1}}$) regardless of the radius. Exceptions are found in minor diffuse components. Post-shock gas located downstream of the cloud is accelerated by pressure gradients along folded magnetic field lines, exhibiting rapid inflow ($V_r<-1\,\mathrm{km\,s^{-1}}$). In addition, gas near the hub shows localized outward motions ($V_r>1\,\mathrm{km\,s^{-1}}$) driven by magnetic tension from the distorted hourglass-shaped field. Despite these local exceptions, the overall kinematic segregation between the dense and diffuse gas agrees with the interpretations of observations for high-speed inflows along filaments in HFSs \citep[e.g.,][]{peretto2014,shimajiri2019,trevio-morales2019}.

Figure~\ref{velocity}~(b) presents the median density-weighted velocity dispersion $\sigma_{v_z}$ along the $z$-axis, corresponding to the shock propagation direction. We find that $\sigma_{v_z}$ tends to increase toward the outer radii, consistent with the observational trends reported by \cite{peretto2023}. The relatively low dispersion near the center suggests that ordered radial inflow dominates over random turbulent motions in the dense hub region. Combined with the radial velocity analysis, these results indicate that, driven by the shock–cloud interaction, mass accretion is selectively channeled through high-density filaments rather than occurring uniformly from the entire cloud, leaving the ambient gas dynamically detached from the primary inflow.

For comparison with observations, Figure~\ref{velocity}~(c1) shows the projected dense-gas distribution and its corresponding line-of-sight velocity field, and Figure~\ref{velocity}~(c2) presents the associated PV diagram. Only gas above a column density threshold of $8\times10^{21}\,\mathrm{cm^{-2}}$, comparable to the mean column density of the identified filaments, is included. The dense radially aligned filaments exhibit a systematic variation of $V_{\mathrm{los,\,z-}}$ across the HFS, with a velocity difference of up to $\sim10\,\mathrm{km\,s^{-1}}$ across the $5\,\mathrm{pc}$ region. This large-scale velocity gradient produces a clear V-shaped pattern in the PV diagram (Figure~\ref{velocity}~(c2)), reflecting the shock-driven reorganization of the velocity field. The overall PV morphology closely resembles the V-shaped signatures reported in high-mass protoclusters and hub–filament systems \citep[e.g.,][]{alvarez2024,salinas2025,sandoval2025}. In contrast to some observed regions that exhibit multiple velocity components, our model produces a single coherent V-shaped feature, consistent with a single shock interaction from one direction.

\subsection{Formation Mechanism of Hub-Filament System}
\begin{figure*}[htbp]
    \gridline{
    \fig{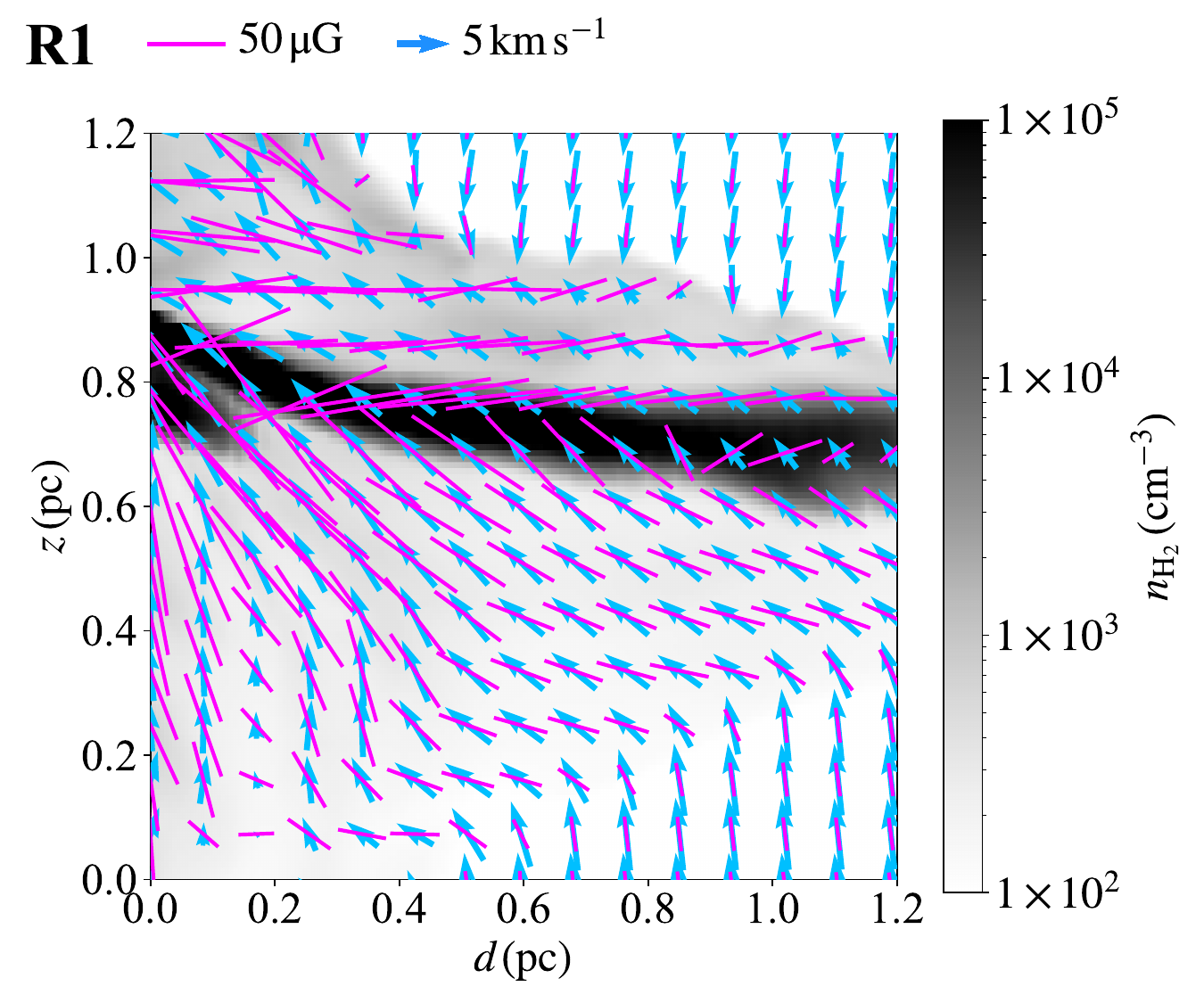}{0.49\textwidth}{}
    \fig{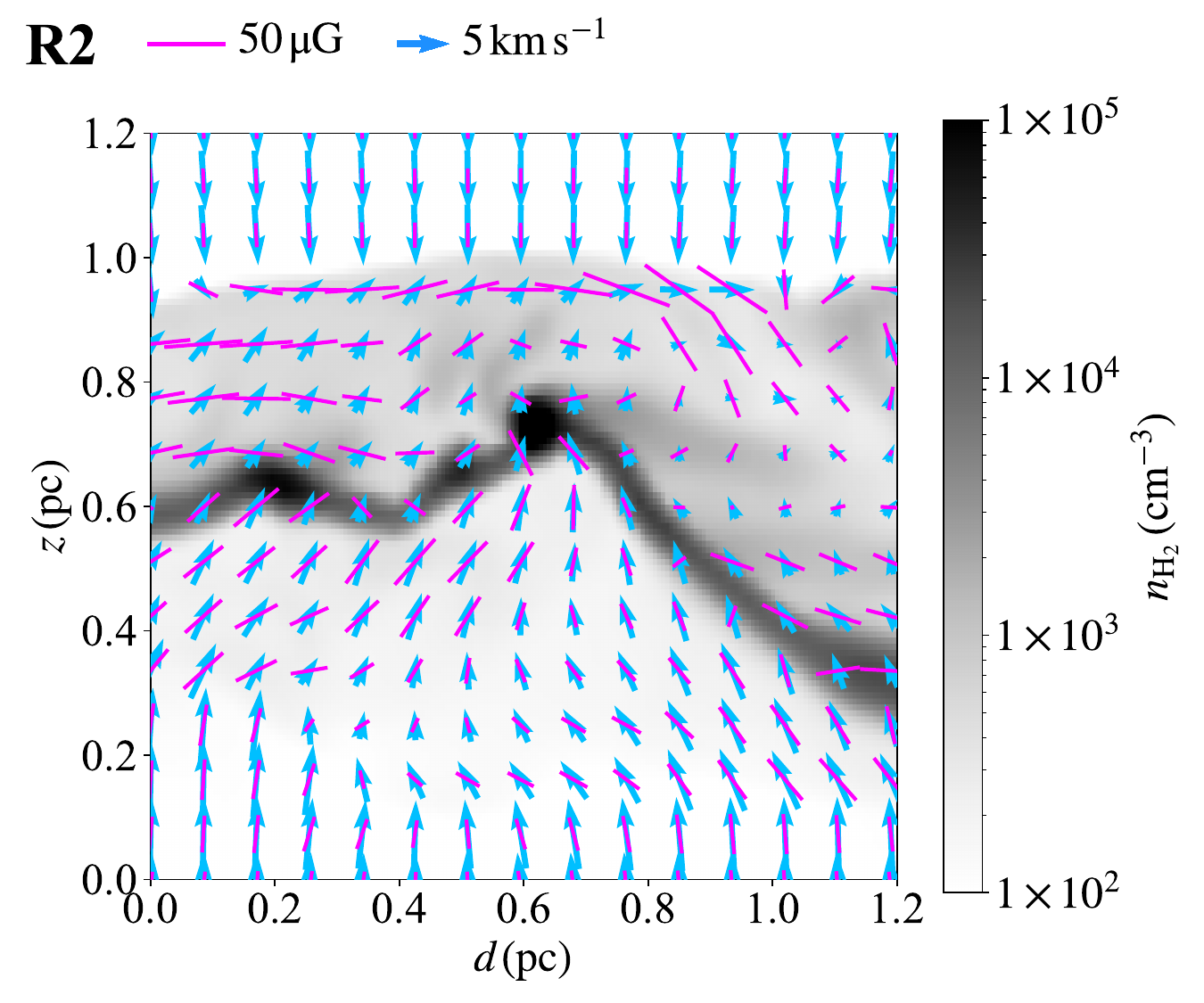}{0.49\textwidth}{}
    }
    \caption{Distributions of density, velocity, and magnetic fields in slices along the filament axis (R1) and across the filament width (R2). The positions of these cuts are indicated by white lines in Figure~\ref{filfinder}~(a). Magenta bars and blue arrows denote the magnetic field and velocity vectors, respectively. The velocity vectors are presented in the shock rest frame ($v_{\mathrm{shock}}=-5\,\mathrm{km\,s^{-1}}$). The grayscale represents the $\mathrm{H_2}$ number density. The horizontal axis $d$ corresponds to the distance along the cut, where the white circle with a gray outline in Figure~\ref{filfinder}~(a) marks the starting points ($d=0$). The vertical axis $z$ represents the vertical position within the slice.}
    \label{v_b}
\end{figure*}

\begin{figure*}[htbp]
    \centering
	\includegraphics[width=\textwidth]{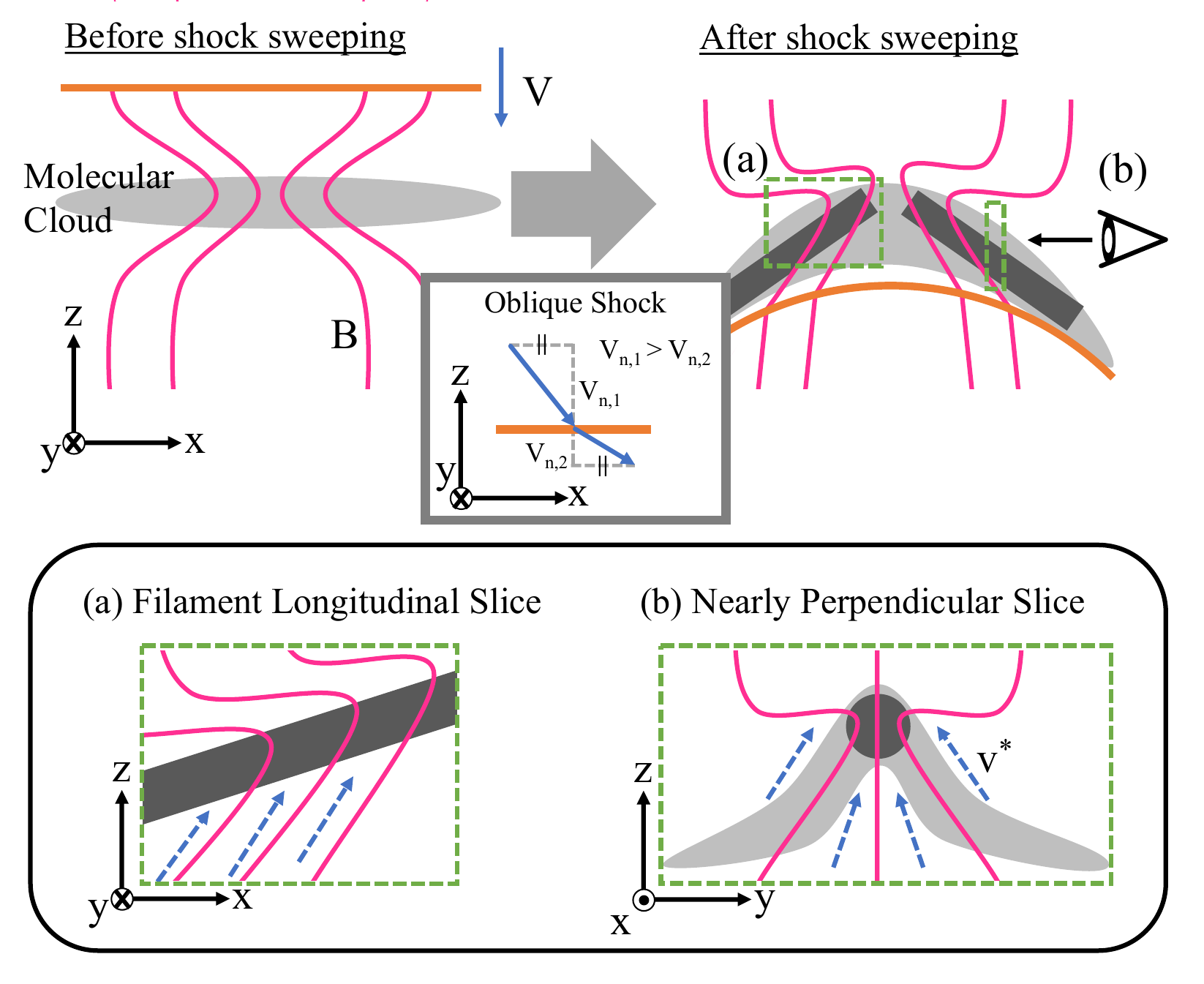} 
    \caption{Schematic illustration of a fast-mode shock propagating through a self-gravitating molecular cloud with hourglass-shaped magnetic field lines. The upper panels show a two-dimensional slice in the $x$--$z$ plane, in which the cloud is flattened along the $z$-direction. The grayscale shading represents the gas density. The blue arrows indicate the velocity field, the magenta curves show the magnetic field lines, and the orange line represents the shock front. (a) Longitudinal slice of the filament, corresponding to cut R1 in Figure~\ref{v_b}. (b) Nearly perpendicular slice of the filament, corresponding to cut R2. 
    The velocity $v^{*}$ denotes the velocity in the shock rest frame.}
    \label{schematic}
\end{figure*}

Figure~\ref{v_b}~(R1) presents the distributions of density, velocity, and magnetic fields in a slice along the axis of a single filament with a width of $\sim0.1\,\mathrm{pc}$.
Across the dense filamentary structure, distinct refraction of magnetic field vectors and field strength enhancement are evident. This signature is characteristic of an oblique fast-mode shock on the hourglass-shaped magnetic field, where shock compression amplifies the tangential magnetic component and locally bends the field lines. In the upstream region ($z\approx0.8\text{--}1.0\,\mathrm{pc}$), the magnetic field is oriented parallel to the filament surface and perpendicular to the velocity vectors in the shock rest frame. In contrast, within the dense filamentary structure and the downstream region ($z\lesssim0.7\,\mathrm{pc}$), the field lines thread the filament and align parallel to the velocity vectors. These features indicate that the interaction between the fast-mode shock and the curved magnetic field redirects the gas flow and channels the post-shock gas along the field lines, forming the radially aligned filament, as schematically illustrated in Figure~\ref{schematic}~(a).

Figure~\ref{v_b}~(R2) presents the distributions of density, velocity, and magnetic fields in a slice across the multiple filaments. The density map reveals a corrugated high-density layer, where the gas density is locally enhanced along the mountain-like ridges in the $z$-direction. These ridges spatially correspond to the filament positions identified in Figure~\ref{filfinder}~(a). This corrugated structure reflects the growth of instabilities resembling the Richtmyer--Meshkov instability, triggered by the shock impact on the weakly turbulent cloud. The magnetic field and velocity structures across this corrugated layer are consistent with those seen in the slice along the filament axis (R1). In the upstream region, the magnetic field is oriented parallel to the corrugated layer and perpendicular to the velocity vectors in the shock rest frame. In contrast, in the downstream region, the field lines thread the layer and align parallel to the velocity vectors, which exhibit a converging flow toward the mountain-like ridges. These features suggest that the gas is not compressed uniformly across the shock plane but is locally accumulated along the corrugations—originating from weak density inhomogeneity—thereby forming the multiple radially aligned filaments (see Figure~\ref{schematic}~(b)).

\section{Discussion}
We show that the interaction of a fast-mode shock with a molecular cloud, characterized by an hourglass-shaped magnetic field bent by self-gravity and density inhomogeneity, forms a radially aligned filament system via a novel formation pathway. This setup, while simplified, corresponds to a physically motivated scenario where a shock (e.g., from a supernova remnant or an expanding H II region) impacts a dense cloud ($n_{\mathrm{H_2}}\sim10^3\text{--}10^4\,\mathrm{cm^{-3}}$; e.g., \citealt{inutsuka2015}). While previous studies have shown that shocks propagating perpendicular to the magnetic field create complex, disordered filamentary density structures \citep{inoue2018}, our results highlight the critical role of the pre-existing field geometry. Even with a shock propagating globally parallel to the background field, the hourglass geometry induces an oblique interaction with the curved field lines. This generates a distinct radially aligned morphology, in contrast to the random structures driven by turbulence. Notably, despite this global morphological difference, the local magnetic field lines still thread the individual filaments, consistent with the local properties reported by \citet{inoue2018} and \citet{abe2021}. We also examined models in which the shock is not initially parallel to the magnetic field and found that radially aligned filaments still develop (Figure~\ref{hf}, right panels), although the overall symmetry becomes weaker with increasing inclination. For an isotropic distribution of shock incidence directions, the probability that a shock arrives within an angle $\psi$ (for $\psi < 90^\circ$) of the magnetic-field axis (counting both $\pm$ directions) is $1-\cos \psi$. For example, this gives $\sim13\%$ for $\psi=30^\circ$, indicating that moderately aligned configurations are not exceedingly rare. These results indicates that the formation of a HFS does not require a perfectly aligned shock–field configuration.

The morphology of the radially aligned filaments formed in our simulation closely resembles the HFSs observed in star-forming regions \citep[e.g.,][]{arzoumanian2022,kumar2020,kumar2022,dewangan2025}. The filament lengths ($1\text{--}3\,\mathrm{pc}$) and mean column densities ($\sim10^{22}\,\mathrm{cm^{-2}}$) are qualitatively consistent with observed values \citep[e.g.,][]{arzoumanian2022,kumar2022}. The measured filament width ($0.06$\text{--}$0.08\,\mathrm{pc}$) is comparable to, though slightly smaller than, the characteristic $\sim0.1\,\mathrm{pc}$ width reported for nearby star-forming filaments \citep[e.g.,][]{arzoumanian2019}. Although HFSs with different spatial scales are also known \citep[e.g.,][]{kumar2022,dewangan2025}, these variations can likely be explained by differences in initial cloud properties (mass and size) and shock parameters (arrival frequency and speed). We note that the mean filament line mass obtained in the present simulation ($\sim22\,\mathrm{\msun\,pc^{-1}}$) is significantly lower than the values reported in massive protocluster HFSs such as G351.77-0.53 and G012.80, where line masses of $\sim10^3$–$10^4\,\mathrm{\msun\,pc^{-1}}$ have been measured \citep[e.g.,][]{reyes2024,salinas2025}. Our model represents either a lower-mass cloud environment compared to massive protocluster HFSs, or an earlier evolutionary stage of such systems. The underlying shock–magnetic interaction mechanism is expected to operate over a broad range of cloud masses and shock strengths, and may extend to systems with higher filament line masses. A comprehensive study on these dependencies, including the effects of cloud geometry and shock incidence angle, will be presented in a forthcoming paper.

Regarding kinematics, Figure~\ref{velocity} shows that high-density gas exhibits larger inward radial velocities toward the hub center, whereas the ambient low-density gas retains low radial velocities. This trend is consistent with the kinematic features of observed HFSs \citep[e.g.,][]{peretto2023,bhadari2025}. While \cite{kumar2022} attributed the filament alignment to the gravitational potential of the hub, our results indicate that high-density filaments become radially aligned and selectively exhibit inflow motions immediately after the shock passage. This selective inflow suggests that the observed alignment is driven primarily by the interaction between the external shock and the magnetic field structure, rather than self-gravity. This kinematic segregation is also reflected in the observationally motivated PV diagram (Figure~\ref{velocity}~(c1) and (c2)), where the dense filamentary components trace a clear V-shaped structure. The overall PV morphology closely resembles the V-shaped signatures reported in recent high-resolution studies \citep[e.g.,][]{alvarez2024,salinas2025,sandoval2025}, indicating that such kinematic patterns can arise from shock–magnetic field interactions without requiring hub-dominated gravitational focusing.

Our model also provides a simple observational prediction for the projected magnetic field morphology in HFSs. In the post-shock dense filamentary region, the magnetic field exhibits refraction and local enhancement due to oblique fast-mode compression of the hourglass-shaped field. As shown in Figure~\ref{v_b}, the field lines thread the filaments and tend to align with the gas flow in the downstream region. The projected orientation of the magnetic field relative to the filament axis depends on the viewing geometry. When viewed from a direction perpendicular to the shock propagation (i.e., along the $x$- or $y$-axis), the projected magnetic field can appear largely perpendicular or oblique to the filament axis (see Figure~\ref{v_b}~(a)). In contrast, when viewed along the shock propagation direction (i.e., along the $z$-axis), the amplified tangential components dominate the plane-of-the-sky projection, and the magnetic field may appear preferentially aligned with the filament. Such projection effects may help explain the coexistence of filament-parallel and filament-perpendicular magnetic field morphologies reported in polarization observations of high-mass star-forming regions \citep[e.g.,][]{sanhueza2021, hwang2022}. A detailed synthetic polarization analysis based on our simulations will be presented in future work.

Finally, we discuss the star formation efficiency (SFE) in the radially aligned filament system. In our simulation, dense gas regions ($n_{\mathrm{H_2}}>1\times10^5\,\mathrm{cm^{-3}}$) are replaced by sink particles, yielding a total sink mass of $1.4\times10^3\,\msun$ at $0.5\,\mathrm{Myr}$ after the shock impact. The total gas mass in the computational domain at this epoch is $1.2\times10^4\,\msun$. Since a continuous inflow is imposed at the boundary ($z=+5\,\mathrm{pc}$), the total mass increases with time by continuous mass supply into the domain. Assuming a core-to-star conversion efficiency of $\sim1/3$ \citep{machida2012}, the SFE at this epoch is estimated to be $4\%$. This value is already consistent with the typical SFEs observed in nearby molecular clouds \citep[e.g.,][]{evans2009,lada2010}, but might be slightly overestimated because of the relatively low density threshold for sink particle creation in Eulerian grid-based simulations. We expect that simulations with much higher spatial resolution and much higher density threshold for sink creation may reduce the total mass of the sink particles. The total filament mass as illustrated in Figure~\ref{filfinder}~(a) is $1.8\times10^3\,\msun$. This mass is computed by integrating the column density over the full filament network identified with \textsc{DisPerSE}, adopting the measured characteristic width, and includes both the prominent central filaments and lower-contrast structures distributed throughout the cloud. The resulting filament mass fraction is $14\%$, comparable to the observed fraction of $10\text{–}20\%$ in nearby molecular clouds \citep{arzoumanian2019}. Considering that $15\%$ of filament mass typically evolves into prestellar cores \citep[e.g.,][]{andre2010,konyves2015} and applying the $1/3$ conversion factor \citep{machida2012}, the potential SFE of the filamentary dense gas is estimated to be roughly $0.7\%$. As shown in Figure~\ref{velocity}, the ambient low-density gas retains significantly lower radial velocities compared to the dense gas. Kinematic segregation limits the rapid mass supply to the central dense region, thereby preventing an excessively high SFE and limiting star formation efficiency.

\section{Summary}
We have investigated the formation of HFSs using three-dimensional ideal MHD simulations of a fast-mode shock propagating into a molecular cloud characterized by an hourglass-shaped magnetic field and density inhomogeneity. Our results are summarized as follows:
\begin{enumerate}
\item The shock-cloud interaction forms a parsec-scale HFS ($1\text{--}3\,\mathrm{pc}$) consisting of multiple filaments radially aligned toward a central hub. The measured filament width is $0.06$\text{--}$0.08\,\mathrm{pc}$, and the corresponding mean line mass ($22\,\msun\,\mathrm{pc^{-1}}$) is slightly above the thermal critical value at $10\,\mathrm{K}$. Statistical analysis confirms a strong radial alignment of the filaments, with deviation angles significantly smaller than those expected for a random orientation. Moderate misalignment between the shock direction and the magnetic-field axis weakens the symmetry but does not suppress the formation of radially aligned filaments.

\item The velocity structure of the system exhibits a clear segregation between dense and diffuse gas. Driven by the shock-cloud interaction, high-density gas within the filaments displays inflow velocities increasing toward the hub center, whereas ambient low-density gas retains low radial velocities. This indicates that mass accretion is channeled through the dense filamentary network.
\item The filament formation is driven by the interaction between the planar shock and the curved magnetic field. The oblique shock generated at the bent field lines amplifies the tangential magnetic component, inducing a magnetically guided inflow. The growth of instabilities resembling Richtmyer--Meshkov modes at the shock interface promotes the fragmentation of the shock-compressed layer into multiple filaments.
\item The estimated SFE is $4\%$, consistent with observations of nearby molecular clouds but might be reduced in future simulations with higher spatial resolution. The potential SFE in filamentary gas is estimated to be only $0.7\%$. These results suggest that the kinematic segregation described above limits the rapid mass supply to the central dense region, thereby preventing an excessively high SFE and naturally regulating star formation.
\end{enumerate}

\begin{acknowledgments} 
We are deeply grateful to Izumi Seno for many insightful discussions throughout this work. We also thank Tomoaki Matsumoto for his major contributions to the code development, and Hajime Fukushima and Masahiro Machida for their helpful comments. Numerical simulations in this paper were conducted by the Cray XC50 (Aterui II) and the Cray XD2000 (Aterui III) at the Center for Computational Astrophysics of National Astronomical Observatory of Japan and the Yukawa-21 at the Yukawa Institute for Theoretical Physics (YITP), Kyoto University. This work was supported in part by MEXT/ JSPS KAKENHI grant Nos. JP25H00394 (S.I.) and JST SPRING, Grant Number JPMJSP2136(S.N.). We acknowledge the use of OpenAI's ChatGPT and Google's Gemini only as grammar checking and editing tools to improve the clarity and readability of the manuscript.
\end{acknowledgments}

\software{DisPerSE \citep{sousbie2011a,sousbie2011b}, Astropy \citep{astropy:2013,astropy:2018,astropy:2022}, Numpy \citep{harris2020array}, Matplotlib \citep{Hunter:2007}}

\bibliography{rf}{}
\bibliographystyle{aasjournalv7}

\end{document}